\documentclass[multphys,vecphys]{svmult}

\usepackage{makeidx}     
\usepackage{graphicx}    
\usepackage{multicol}    

\makeindex             

\begin{document}

\title*{The Galactic center interstellar medium as seen by ISO}
\author{N. J. Rodriguez-Fernandez
\inst{1}
\and
J. Martin-Pintado\inst{2}
}
\institute{LERMA (UMR 8112), Observatoire de Paris / CNRS, 61 Av. de l'Observatoire, 75014 Paris, France
\texttt{nemesio.rodriguez@obspm.fr}
\and DAMIR, IEM, CSIC, Serrano 121,  Madrid, Spain
}
%
%
\maketitle

This paper deals with the heating and the ionization of the interstellar
medium (ISM) in the 500 central pc of the Milky Way (hereafter Galactic center, GC).
We review the results of {\it Infrared Space Observatory} (ISO) observations of a sample of GC molecular clouds located far from thermal radiocontinuum or far-infrared sources.
For the first time, we have been able to study in detail the
dust continuum spectra from 40 to 190 $\mu$m founding a warm (30-40 K) dust component in addition to the well known 15-20 K  component.
Fine-structure lines observations have revealed the presence of diffuse ionized gas associated with the molecular clouds. The effective temperature of the ionizing radiation is higher than 33000 K.
ISO has also allow us to measure the fraction of
warm ($\sim$ 150 K) H$_2$ in the GC clouds, which is on average of 30 $\%$.
The observations of the warm (a few 100 K) neutral gas are compatible with a Photon Dominated Region (PDR) scenario.

\section{Dust}
\begin{figure}
\centering
\includegraphics[width=12cm]{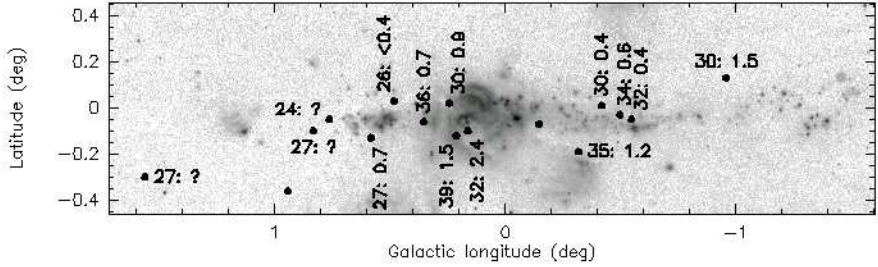}
\caption{Sources positions (black dots) overlaid on the 20 $\mu$m image
by MSX. For each source numbers separated by a colon represent the
temperature of the warm dust component in K and the [O III] 88 to
 [N II] 122 $\mu$m flux ratio }
\label{fig1}
\end{figure}

We have been able to study the full spectrum of the dust continuum
emission from 40 to 190 $\mu$m.
Two gray bodies are needed to model the spectra: a cold component
with  similar temperature for all the sources of 15-18 K and a warmer
component with a temperature of 26 K to 39 K depending on the source
(see Rodriguez-Fernandez et al. 2004).
The cold dust is a well known component of the GC ISM from previous
far-infrared and sub-millimeter studies.
On the contrary, this is the first time we observe the warm
dust component.
As shown in Fig. \ref{fig1},
the warm component temperature and luminosity are well correlated with
the 20 $\mu$m intensity in the images taken by
 {\it Midcourse Space Experiment} (MSX).

\section{Ionized gas}
\begin{figure}
\centerline{
\includegraphics[width=5.5cm]{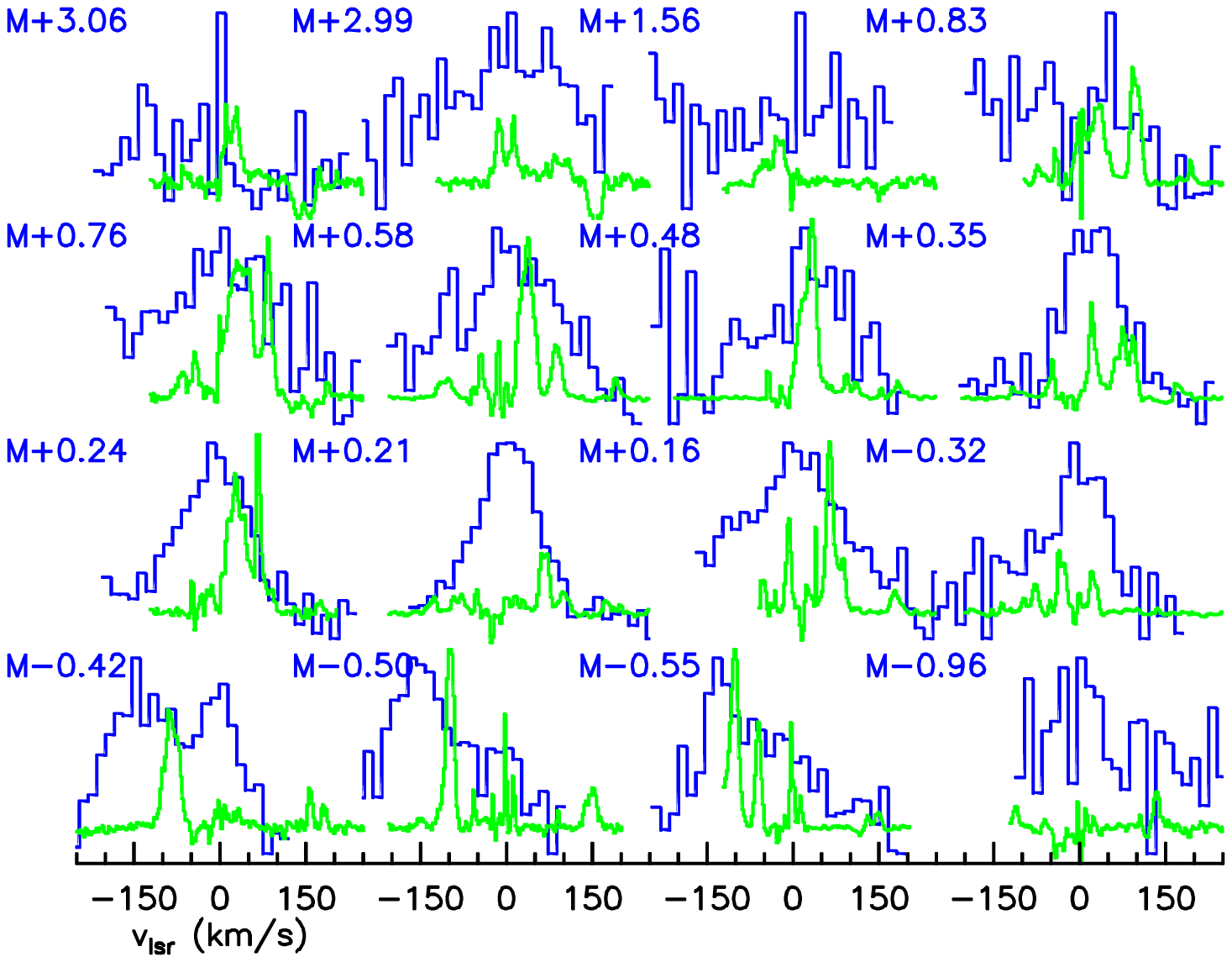}
\includegraphics[width=6cm]{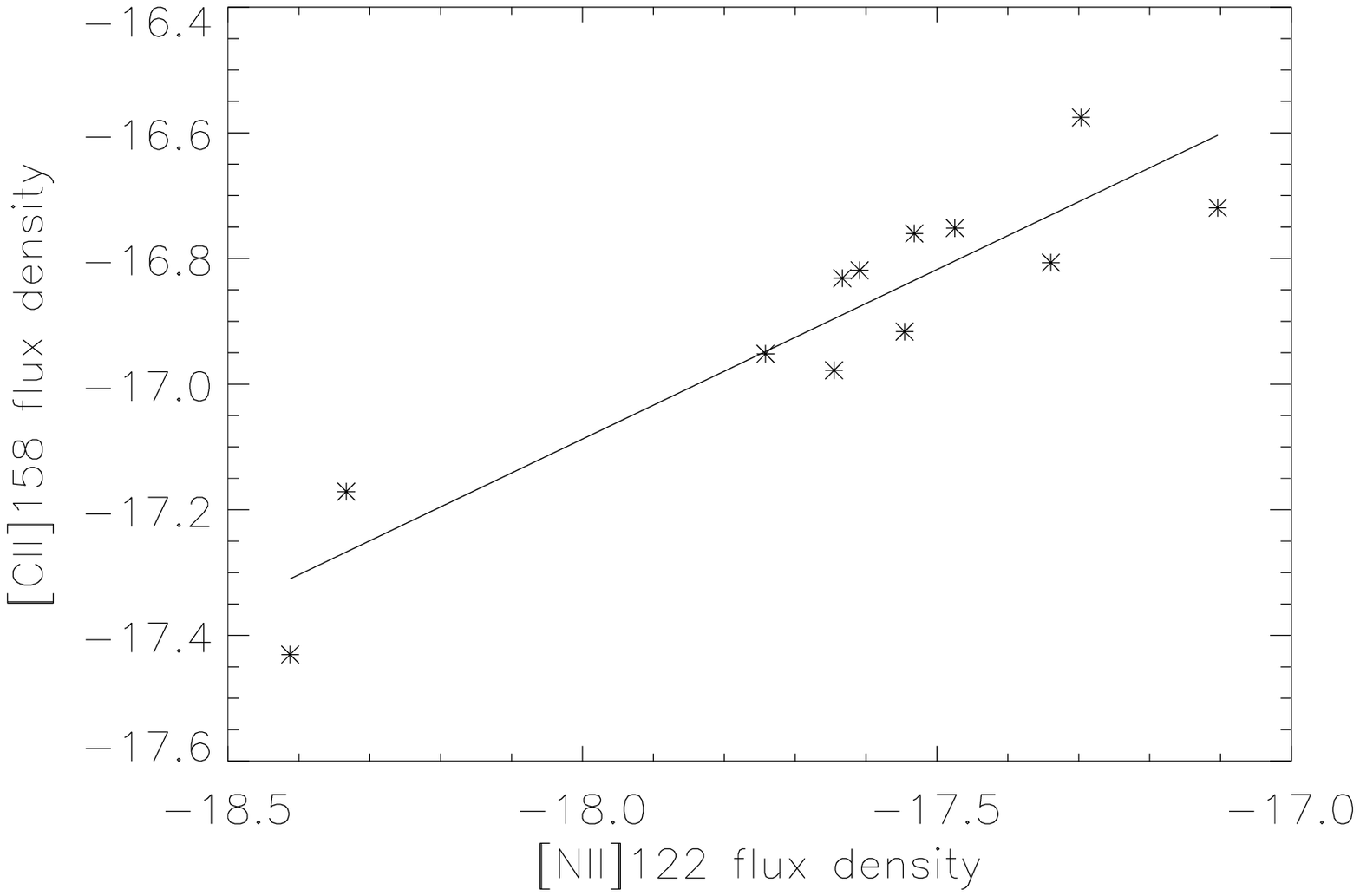}
}
\caption{Left panel: comparison of the [N II] 122 $\mu$m and the $^{13}$CO(1-0)
line profiles.
Right panel: [C II] 158 $\mu$m versus [N II] 122 $\mu$m fluxes and least squares fit.}
\label{fig2}
\end{figure}

We have detected fine structure lines of ionized species as [C II] 158 $\mu$m
and [Si I] 34 $\mu$m in all the sources. In most of them we have also detected
lines like [N II] 122 $\mu$m, [Ne II] 13 $\mu$m or [S III] 34 $\mu$m.
In 11 of the 18 sources we have even detected the [O III] 88 $\mu$m
(with excitational potential of 35 eV).
(Rodriguez-Fernandez et al. 2001b; Rodriguez-Fernandez \& Martin-Pintado 2004).
The lines observed with the LWS Fabry-Perot (velocity resolution
of $\sim 30$\,km\,s$^{-1}$) are very broad, with linewidths up to 150 km\,s$^{-1}$.
Left panel of Fig. \ref{fig2} shows the [N II] spectra overplotted on the
$^{13}$CO(1-0) data of Rodriguez-Fernandez et al. (2001a).
Taking into account the rather different spectral
resolutions, for most of the sources the Fabry-Perot lines profiles are similar to those of the $^{13}$CO(1-0) lines.
[S III] and [O III] lines ratios imply that the electron densities are
lower than a few 10 cm$^{-3}$.
Therefore, line profiles and electron densities are consistent with the ionized
gas arising in the low density envelopes of the molecular clouds.
The  good correlation between the the [C~II] 158 $\mu$m and the [N~II] 122 $\mu$m fluxes (right panel of Fig. \ref{fig2})  implies that
part of the [C~II] flux also arises the ionized gas component (Heiles 1994).

Figure \ref{fig1} shows one of the lines ratios sensitive to the effective temperature
of the ionizing radiation
that can be derived for most of the sources ([O III] 88/[N II] 122) over the 20 $\mu$m MSX image. The higher lines ratios  are measured in those sources with the higher dust temperature (Section 1)
which are those located in the surrounding of the 20 $\mu$m emission features.
The association of warm dust and ionized gas imply heating by stars radiation.
The ionizing sources are probably located in the vicinity of the 20 $\mu$m peaks
but the radiation reach large distances due to the inhomogeneity of the medium,
i.e., the general scenario is probably similar to that found by Rodriguez-Fernandez et al. (2001b) in the Radio Arc region.
CLOUDY models show that the effective temperature of the radiation is
 higher than 33000 K (Rodriguez-Fernandez $\&$ Martin-Pintado 2004).

\section{Warm neutral gas}
Observations of H$_2$ pure-rotational lines have allow us to
measure the total column density of warm gas in the GC clouds.
The column density of warm ($\sim$ 150 K) H$_2$ is $\sim 10^{22}$ cm$^{-2}$.
In addition there is a $\sim$ 500 K gas component with a column density lower
than 1$\%$ of that at 150 K.
Comparing with the H$_2$ column density derived from CO we see that the fraction of
warm H$_2$ is on average of 30 $\%$ but it is as high as $\sim 100 \%$ for some of the
sources.
The H$_2$ line intensities can be reproduced both by shocks or PDRs models.
In particular, the observed temperature gradient is in perfect agreement with both
model predictions and ISO observations of well-known PDRs
(Rodriguez-Fernandez et al. 2001a)

In addition, we have observed the warm (a few 100 K) neutral medium by means of
fine structure lines of species like O, C$^+$ or Si$^+$, which are the
main cooling lines in these medium.
The lines to continuum ratio ($\sim 0.3 \%$) is typical of PDRs.
Indeed, the discovery of ionized gas associated with the molecular clouds
implies that PDRs are expected in the interface between the ionized and
the molecular gas.
From the PDRs diagnostic diagrams of Wolfire et al. (1996) we estimate
a density of 10$^3$ cm$^{-3}$ and an incident field 10$^3$ times higher
than the local interstellar radiation field.

We have also found that the ratio of warm to cold H$_2$ increases with
an increasing fine-structure lines to far-infrared continuum ratio.
This fact points to
a common origin for the warm H$_2$ and the fine structure lines.
(Rodriguez-Fernandez et al. 2004).

\section{Discussion and conclusions}
The heating of the GC clouds is a long-standing problem.
The discrepancy of the dust and gas temperatures and the apparently
lack of ionized gas associated to the molecular clouds are usually invoked
to rule out radiative heating mechanisms.
However,
the picture arising from the ISO observations is a complex medium
where the molecular clouds are irradiated from the exterior by hot
radiation arising from relatively hot and distant sources.
The surfaces of the clouds are ionized and
there is a warm (150 K) neutral gas component
in the interface between the ionized and the molecular gas.
The 30-40 K dust component should be associated with the 150 K gas
component and both heated in PDRs.
It is noteworthy that the discrepancy of gas and dust temperatures only
rules out gas heating by collisions with hot dust but it does not rule
out {\it any} radiative heating mechanisms as photoelectric effect on the dust grains in a PDR. Indeed a gas temperature of 150 K and a dust temperature of
$\sim35$ K as measured in the GC clouds it is exactly what it is expected in
a 10$^3$ cm$^{-3}$ and G$_0$=10$^3$ PDR (Hollenbach et al. 1991).

On the other hand,
the extended distribution and high abundances of fragile molecules
when iluminated by ultra-violet radiation (see for instance Martin-Pintado
et al. 2001) and the large line-widths are the best probe of the presence
of turbulence or low velocity shocks in the GC clouds.
Shocks, turbulence and PDRs probably coexist in the GC.
Unfortunately, it is still not clear  how much gas is heated in PDRs and how much is heated by other mechanisms.
In any case, the new ISO data show that the importance of radiation in the heating of the GC clouds should be revisited and that a significant fraction of the warm gas in the
GC clouds arise in standard PDRs.

\vspace{5mm}
{\bf Acknowledgments:}
The authors acknowledge the collaboration of  A. Fuente, P. de Vicente, T.L.Wilson and S. Huttemeister in some of the work presented in this contribution.
N.J.R-F acknowledges fruitful discussions with J.R. Goicoechea and  M. Morris during the conference and support by a Marie Curie Fellowship of the European Community
program``Improving Human Research Potential and the Socio-economic Knowledge base''
under contract number HPMF-CT-2002-01677.
J. M-P acknowledges {\it Ministerio de Ciencia y Tecnologia} grants ESP2002-01627 and AYA2002-10113-E.

%

\printindex
\end{document}